\documentstyle[editedvolume]{crckapb} 

\begin{opening}
\title{``E+A'' GALAXIES:  ENVIRONMENT AND EVOLUTION}
\author{Ann I. Zabludoff}
\institute{UCO/Lick Observatory and University of California,\\
Santa Cruz, CA, USA}

\end{opening}

\runningtitle{``E+A'' GALAXIES:  ENVIRONMENT AND EVOLUTION}

\begin{document}

\section{Introduction}
\renewcommand{\thefootnote}{\fnsymbol{footnote}}
One important approach to the study of galaxy evolution is to identify
those galaxies whose spectral and/or morphological characteristics suggest
that they are in transition.  For example, ``E+A'' galaxies\footnote[1]{
The term ``E+A'' is a bit of a misnomer.  The Mg, Fe, and Ca lines
observed in the spectra of these galaxies are consistent with the
stellar populations of ellipticals or ``E''s.
The additional ``A'' designation arose from the galaxies' strong
Balmer absorption lines, which are characteristic of A stars.
Because the morphologies of ``E+A''s now appear to range from
spheroidals to disks, a more apt, and exclusively spectroscopic, designation
is ``K+A'' (cf. Franx 1993).  Nevertheless, we use
``E+A'' throughout this paper for 
historical reasons.}\renewcommand{\thefootnote}{\arabic{footnote}}
, which have
strong Balmer absorption lines and no significant [OII] emission, are generally
interpreted as post-starburst galaxies in which the star formation ceased
within the last $\sim$ Gyr (Figure 1).  This transition between
a star forming and non-star forming state is a critical link in any
galaxy evolution model in which a blue, star forming disk galaxy
evolves into a S0 or elliptical.  
Another possible evolutionary track is that the 
star formation in an ``E+A'' resumes at some later time, if enough gas
remains in the galaxy after its starburst ends.
Given this ambiguity, it is important to investigate (1) the
environment's role 
in ``E+A'' evolution, (2) the stellar and gas morphologies of ``E+A''s,
(3) the likely progenitors of ``E+A''s, and (4) how common the
``E+A'' phase is in the evolution of galaxies.

This proceeding summarizes recent results from 
several inter-related projects designed to address these questions.
These projects focus on a sample of 21 nearby
``E+A'' galaxies ($0.05 < z < 0.15$;
Zabludoff et al. 1996) drawn from the Las Campanas Redshift 
Survey (Shectman et al. 1996).  These studies include VLA
and HST observations, in addition to comparisons of these
data with galaxy-galaxy interaction
simulations and stellar population synthesis models.
My collaborators are D. Zaritsky (UCO/Lick), J. van Gorkom (Columbia), 
C. Mihos (Case Western),
I. Smail (Durham), G. Bruzual (CIDA), S. Charlot (IAP), M. Franx
(Leiden), and R. Bernstein (OCIW).

\begin{figure}
\includegraphics{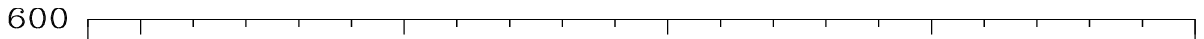}
\vspace{8cm}
\caption{Identification of lines in the
rest frame spectrum of an ``E+A'' galaxy, which is dominated
by a young ``A'' stellar component.  
The residual sky line at 5577 \AA\ has been excised.
Note the absence of [O II] emission.}
\end{figure}

\section{Environment and ``E+A'' Evolution}
The role of environment in the evolution of ``E+A'' galaxies, specifically
in producing the initial burst of star formation, in ending it, and in
allowing it to resume, is unknown.  In past work, the 
detection of ``E+A''s almost
exclusively in distant clusters led to speculation that
these galaxies represent an evolutionary sequence unique to or
most efficient in cluster environments.  
The existence of such
a cluster-dependent evolutionary sequence would suggest that
the cluster environment, in the form of the intra-cluster medium,
galaxy harassment, or the global potential,
is responsible for the recent star formation history of ``E+A''s
and, by extension, for the Butcher-Oemler effect (Butcher \& Oemler 1978) 
in clusters.  
In contrast to this line of reasoning, Schweizer (1982, 1996) and others
find several nearby ``E+A''s that appear to lie outside
the hot, dense environments of clusters and that have highly disturbed
morphologies consistent with the products of galaxy-galaxy mergers.
To isolate at least one mechanism that governs 
the evolution of ``E+A'' galaxies
requires a statistical inventory of the environments in which ``E+A''s form.
The Las Campanas Redshift Survey (LCRS), which includes high
signal-to-noise spectra for $\sim 11000$ galaxies with $0.05 < z < 0.15$,
is the ideal sample with which to characterize the environments of ``E+A''s. 

\begin{figure}
\includegraphics{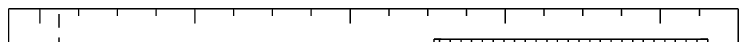}
\vspace{8cm}
\caption{Plot of average Balmer line
absorption $\langle H \rangle$ {\it vs.} [O II] line emission
EW[O II] for the 11113 LCRS galaxies
with $S/N > 8$ and $0.05 < z < 0.13$.  
The dashed line encloses the region,
$\langle H \rangle > 5.5$ \AA\ and EW[O II] $ < 2.5$ \AA, from which the
sample of 21 ``E+A'' galaxies (large points) is drawn.
The inset shows that EW[O II] cut excludes
galaxies with a more than 2$\sigma$ detection.}
\end{figure}

To identify ``E+A'' galaxies in the LCRS having properties consistent with
those of known ``E+A''s, we plot
the distribution of Balmer absorption line and [OII] emission
line strength for the LCRS galaxies (Figure 2; Zabludoff et al. 1996).  
The ``E+A''s are selected to have the strongest
Balmer absorption lines (the average of the equivalent widths of H$\beta$,
$\gamma$, $\delta$ is $> 5.5$ \AA) and weakest [O II] emission-line
equivalent widths ($< 2.5$ \AA, which corresponds to a detection of [O
II] of less than 2$\sigma$ significance) of any of the galaxies in the
survey.  We test whether these 21 ``E+A''s lie 
in rich clusters in several ways,
including calculating the local galaxy density around each ``E+A'' and also
checking whether the ``E+A'' is a member of a rich cluster
in the LCRS group catalog (Tucker 1994).  Surprisingly,
a large fraction ($\sim 75\%$) of nearby
``E+A''s lie in the field, well outside of clusters and rich groups of
galaxies.  We conclude that interactions with the cluster
environment are not essential for ``E+A'' formation and therefore that
the presence of these galaxies in distant clusters does not provide
strong evidence for the effects of cluster environment on galaxy
evolution.

If one mechanism is responsible for ``E+A'' formation, then the
observations that ``E+A''s exist in the field
and that at least five of the 21 in our
sample have clear tidal features argue that
galaxy-galaxy interactions and mergers are that mechanism.  
The most likely environments for such mergers are poor groups of
galaxies, which have lower velocity dispersions than clusters and
higher galaxy densities than the field.  Groups are correlated with rich
clusters and, in hierarchical models, fall into clusters in
greater numbers at intermediate redshifts than they do today
(cf. Bower 1991; Lacey \& Cole 1993; Kauffmann 1994).  
When combined with the strong evolution observed in the
field population (cf. Broadhurst et al. 1988; Lilly et al. 1995), 
our work suggests that the Butcher-Oemler effect may reflect the
typical evolution of galaxies in groups and in the field 
rather than the influence of
clusters on the star formation history of galaxies.

\section{Stellar and HI Morphologies of ``E+A''s}
Is the transition between galaxy types implied
by the post-star formation spectrum of an ``E+A'' seen in its morphology?
The two HST images that we have obtained to date suggest a
morphological transition.  One ``E+A'' has an E type morphology,
but has extended tidal tails.  The other ``E+A'' is a barred S0.
If star formation does not resume in these galaxies, they
will look like early types after their blue stars die.
One interesting unanswered question is why galaxies of such different
morphologies have such similar spectra.

There is substantial evidence that
galaxy-galaxy interactions
increase star formation rates.  While the effects of such interactions are
consistent with the starburst history of ``E+A''s, the mechanism by
which the star formation stops is still a mystery.  The HI morphologies
of ``E+A''s can provide some clues.

In the first LCRS ``E+A'' for which we have obtained VLA data (Figure 3), 
there are
clear HI tidal tails similar to those of the Antennae (Hibbard \& Mihos 1995).
These tails support the galaxy-galaxy interaction picture for
``E+A'' formation.  Perhaps even more interesting is the distribution
of the gas not in the tails.  The gas mass is comparable
to that in disk galaxies ($\sim 5 \times 10^9$ M$_{\odot}$), but it 
is extended over $50h^{-1}$ kpc.  Thus, the lack of star formation in
this ``E+A'' is not due to an absence of gas, but perhaps to the low density
of that gas.  It is possible that this gas will someday fall back into the
galaxy and generate new star formation.

The rarefied HI gas in this ``E+A'' suggests that the subsequent
evolution of such galaxies could be affected by environment.
Extended gas is easier to strip by ram pressure than
that in galactic disks.
Therefore, if this ``E+A'' formed in a subcluster, instead of the field,
it is likely that the effects of the intracluster medium would preclude
subsequent star formation.
While derived from only one VLA observation to date, this speculation
may illuminate one source of the difference between the
galaxy morphologies in clusters and in the field.  In this spirit, our current
observational program is to compare the gas distributions in cluster
and field ``E+A''s.

\begin{figure}
\includegraphics{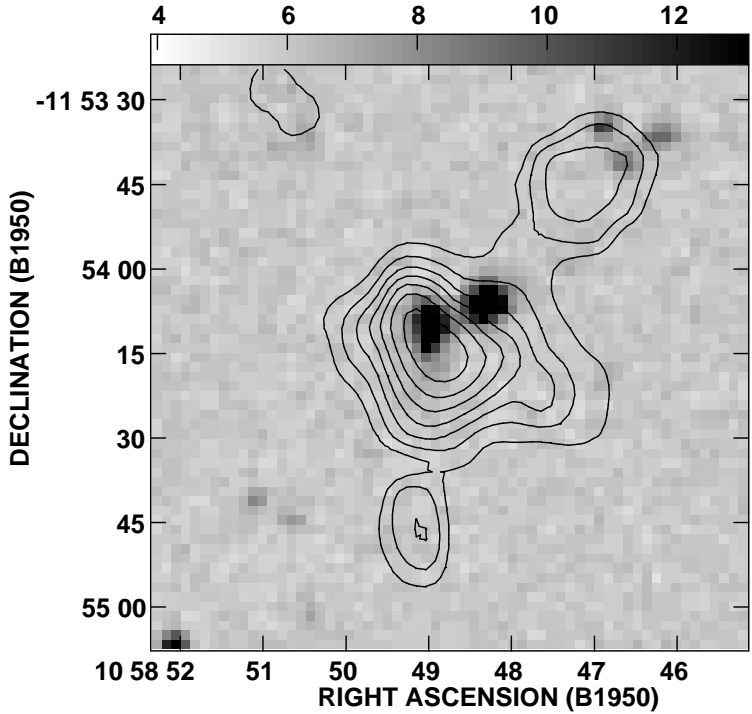}
\vspace{10cm}
\caption{The HI distribution superposed on the Scanned Digitized
Sky Survey SERC $b_J$ image of one of the bluest ``E+A''s in the sample.
The central
HI gas has a mass of $\sim 5 \times 10^9$M$_{\odot}$, consistent with
that in late type galaxies, but it is extended over
more than $50h^{-1}$ kpc.}
\end{figure}

\section{``E+A'' Progenitors}
As discussed above,
the clear tidal features in some ``E+A''s indicate that these galaxies
have evolved morphologically, in addition to spectroscopically.
To identify the morphologies of the most likely ``E+A'' progenitors,
we assume that two progenitors merge to form an ``E+A'' and
derive limits on their gas-to-stellar masses from the strength of
the ``E+A'' starburst (as inferred from a comparison of the
Balmer absorption lines and the 4000\AA\ break strengths with
stellar population
synthesis models; Bruzual \& Charlot 1995).  If the gas-to-stellar masses 
of the ``E+A'' progenitors are consistent with those of gas-rich,
disk galaxies, and a particular ``E+A'' is an S0 or E,
then we can conclude that a morphological transformation has occurred.

For most of the 21 LCRS ``E+A''s, the (HI+H$_2$)-to-stellar mass ratios
of a pair of Sa-c spirals provide sufficient gas to
generate burst strengths corresponding to 10-30\% of the total stellar mass
in the ``E+A''.  Note that we assume a standard Scalo IMF and a
star formation efficiency ({\it i.e.,} fraction of gas converted to stars) of
50\% or less.  
However, for the bluest three ``E+A''s in the sample ({\it e.g.,} 
Figure 1), the
burst strengths of $\sim 50\%$ cannot be reproduced without relaxing some of 
these assumptions.  For example, either both merging progenitors are
late Sd disks, or at least one is a low surface brightness, Malin I
type galaxy, or the star formation efficiency of the resulting burst is an
extraordinary 100\% (in contrast with the $\sim 50\%$ efficiencies 
of the brightest IRAS ultra-luminous galaxies).
Although based on stellar population synthesis models that are still
incomplete,
these results support the picture that ``E+A''s are a phase
of galaxy evolution in which blue, star-forming disk galaxies are transformed
via galaxy-galaxy encounters into early type S0 and E galaxies.

\section{How Common is the ``E+A'' Phase?}
Are ``E+A''s rare objects or do they represent a short-lived phase in the
evolution of many galaxies?  From a comparison of the 21 ``E+A''
spectra with stellar population
synthesis models, we estimate that the duration of the ``E+A'' phase
is $< 0.8$ Gyr.  The fraction of galaxies that are ``E+A''s in the nearby
universe is $21/11113 = 0.002$.  Therefore, at least 4\% of galaxies
could have passed through an ``E+A'' phase within a Hubble time,
a fraction which would constitute
a significant number of the early types in the field.
We plan to improve this estimate by comparing the HST images, HI maps,
and follow-up long-slit spectra with simulations of galaxy-galaxy interactions
(cf. Mihos \& Hernquist 1994).  
The internal kinematics and morphological features
on small scales should better constrain the time elapsed since the
starburst ended and thus the duration of the ``E+A'' phase in galaxies.

\end{document}